\documentclass[twocolumn,showpacs,preprintnumbers]{revtex4}
\usepackage{graphicx}
\usepackage{dcolumn}
\usepackage{bm}

\begin{document}

\title{Gravitational time advancement and its possible detection}
\author{ A. Bhadra}
\email{aru_bhadra@yahoo.com}
\affiliation{High Energy \& Cosmic Ray Research Centre, University of North Bengal,
Siliguri, WB 734013 INDIA }
\author{K. K. Nandi}
\email{kamalnandi1952@yahoo.co.in}
\affiliation{Department of Mathematics, University of North Bengal, Siliguri, WB 734013
INDIA }

\begin{abstract}
The gravitational time advancement is a natural but a consequence of curve space-time geometry. In the present work the expressions of gravitational time advancement have been obtained for geodesic motions. The situation when the distance of signal travel is small in comparison to the distance of closest approach has also been considered. The possibility of experimental detection of time advancement effect has been explored.
\end{abstract}

\pacs{03.30.+p,04.80.Cc }
\keywords{relativity, negative time delay, detection}
\maketitle

\section{Introduction}

The Shapiro time delay (also known as gravitational time delay) [1] constitutes one of the three classic solar system tests of general relativity. The effect arises from both the spatial and temporal coefficients of space-time metric and thus serves as a comprehensive test of general relativity. Nowadays the effect has also been employed as a tool to extract information about the distribution of matter in the Universe, particularly to detect dark matter in our Galaxy.

The general perception about the Shapiro effect is that due to the influence of a gravitating object the average speed of light decreases from its canonical special relativistic value $c_{0}$ and hence the signal always suffers an additional (positive) non-Newtonian delay. We wish to point out that this is \textit{not} the case in general; depending on the position of
the observer, the delay can as well be negative implying a time advancement. Note that the effect of time advancement does not violate causality, information could not be sent by an observer into his/her own past exploiting the effect, neither can it be used for a warp drive or time machine. The reason is that we are not considering motion in a configuration with matter violating known energy conditions.

So far all conclusive gravitational time delay measurements, including the one using Cassini spacecraft that has verified gravity with a remarkable accuracy of about $2.3$ parts in $10^{5}$ [2], have tested general relativity in the gravitational field of the Sun and in all such cases observers were far away from the gravitating object. Consequently the delay
has been found to be positive in all such measurements, as expected [3]. Instead if we consider the situation that light signal is sent say from the Earth's surface to a ceratin distance from where the signal is reflected back to the point of transmission, then the observer should notice a time advancement (see the following section). Here we explore the
possibility of detecting the gravitational time advancement effect in a future astrometric experiment. We particularly consider the situation  in which light signal will be sent from Earth to one of its artificial satellites/space station from where the signal will be reflected back along the same trajectory and the total travel time between signal transmission and reception will be measured with required precision. When a light signal is sent from Earth to one of its artificial satelites, the signal would also come under the influence of Sun's gravity and one needs to isolate time advancement effect due solely to Earth's gravity from the resulting motion. For the purpose we obtain expression for the magnitude of the time advancement/delay effect when the distance between the points of transmission and reflection of the signal is very small in comparison to the impact parameter.

The organization of the article is as follows. In the next section we describe the gravitational time advancement. In section 3 we obtain the model independent expressions for time advancement/delay when the length of signal propagation is very small. In section 4 we explore the possibility of detecting time advancement effect experimentally. Finally we conclude our results in section 5.

\section{Gravitational time advancement}

The proposed effect can be understood by considering the following scenario: A radar signal is sent from the surface of the Earth ($A$) to a point $B$ close to the Sun, (B is the point of closest approach for the trajectory) as in the Fig.1 from where the signal is reflected back along its original trajectory to the Earth. 

\begin{figure}
\centering \includegraphics[width=0.45\textwidth,clip]{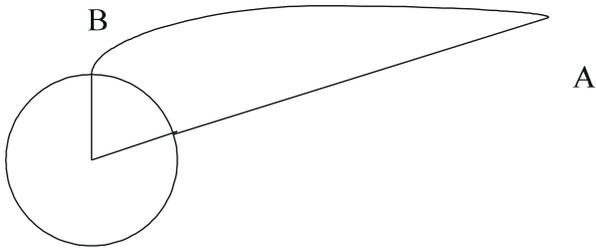} \hfill
\caption{Schematic view of gravitational time delay/advancement}
\end{figure}

Assuming standard Schwarzschild geometry, to first order $\mu \equiv GM/c_{0}^{2}$, the well-known coordinate time delay in round trip journey from Earth $A$ to the point $B$ and back is given by [1,4]
\begin{eqnarray}
c_{0}\Delta t_{AB} &=&2\sqrt{r_{A}^{2}-r_{B}^{2}}+4\mu _{\odot }\ln \frac{
r_{A}+\sqrt{r_{A}^{2}-r_{B}^{2}}}{r_{B}}+  \nonumber \\
&&2\mu _{\odot }\left( \frac{r_{A}-r_{B}}{r_{A}+r_{B}}\right) ^{1/2}\;,
\end{eqnarray}

where $\mu _{\odot }\equiv GM_{\odot }/c_{0}^{2}$, $M_{\odot }$ being the mass of the Sun and $r_{A}$ and $r_{B}$ are the values of coordinate $r$ evaluated at the positions of $A$ and $B$ respectively. The difference in proper time between transmission and reception of the signal to be measured by the observer at $A$ is

\begin{eqnarray}
c_{0}\Delta \tau _{AB} &\simeq &\left( 1-\mu _{\odot }/r_{A}\right) \Delta
t_{AB}  \nonumber \\
&\simeq &2\sqrt{r_{A}^{2}-r_{B}^{2}}+4\mu _{\odot }\ln \frac{r_{A}+\sqrt{
r_{A}^{2}-r_{B}^{2}}}{r_{B}}+  \nonumber \\
&&2\mu _{\odot }\left( \frac{r_{A}-r_{B}}{r_{A}+r_{B}}\right) ^{1/2}-2\mu
_{\odot }\frac{\sqrt{r_{A}^{2}-r_{B}^{2}}}{r_{A}}\;.
\end{eqnarray}

Hence the signal takes an excess time over the time that it would have taken in the absence of the Sun and the delay (the part proportional to $\mu _{\odot })$ is positive for any $r_{A}$ (see fig 2(a)).

\begin{figure}[h]
\centering  \includegraphics[width=0.45\textwidth,clip]{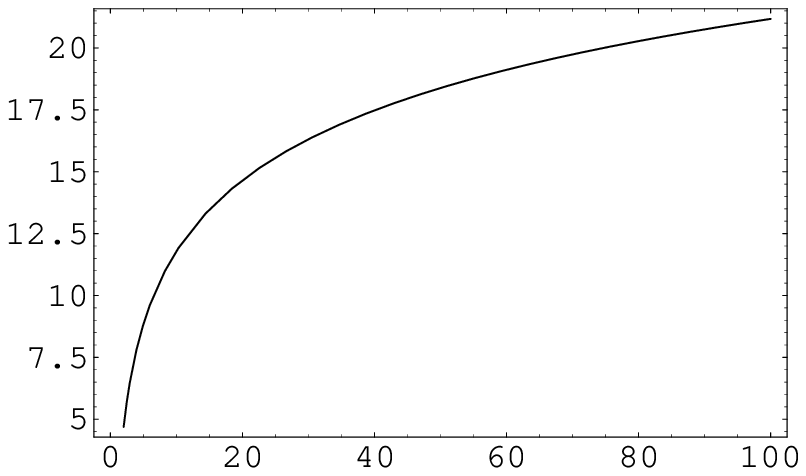}
\caption{Gravitational time delay (along y-axis) as a function of ratio
between $r_{A}$ and $r_{B}$ (along x-axis) when the observer is at $A$. The
delay has been given in artitrary units.}
\end{figure}

However, if the observer is at the point $B$ instead of $A$, the coordinate time delay for the round trip journey from the position $B$ to $A$ and again back to $B$ would remain the same as given in Eq.(1) but the difference in proper time to be measured by the observer at $B$ now reads 

\begin{eqnarray}
c_{0}\Delta \tau _{AB} &\simeq &\left( 1-\mu _{\odot }/r_{B}\right) \Delta
t_{AB}  \nonumber \\
&\simeq &2\sqrt{r_{A}^{2}-r_{B}^{2}}+4\mu _{\odot } \ln \frac{r_{A}+\sqrt{
r_{A}^{2}-r_{B}^{2}}}{r_{B}} +  \nonumber \\
&&2\mu _{\odot }\left( \frac{r_{A}-r_{B}}{r_{A}+r_{B}}\right) ^{1/2} -2\mu
_{\odot }\frac{\sqrt{r_{A}^{2}-r_{B}^{2}}}{r_{B}}\;.
\end{eqnarray}

Because of the last term in the right hand side of Eq.(3), the delay works out to be negative as clearly revealed from the fig 2(b).

\begin{figure}[h]
\centering \includegraphics[width=0.45\textwidth,clip]{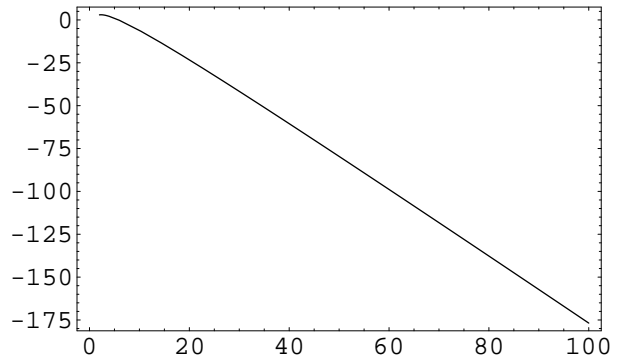}
\caption{Gravitational time delay (along y-axis) as a function of ratio
between $r_{A}$ and $r_{B}$ (along x-axis) when the observer is at $B$. The
delay has been given in arbitrary units.}
\end{figure}

\section{Gravitational time advancement/delay for small distance travel}

As mentioned earlier, in all gravitational time delay measurements conducted so far observers were far away from the gravitating object and the distances between the points of signal transmission and reflection were much larger in
comparison to the distance of closest approach. In situations where the signal travel distance is small relative to the distance of closest approach, the standard expression for time advancement as given by Eq.(3) needs to be applied with caution, giving due importance to the operational meaning of the distance of light propagation. In the following we would obtain \textit{model independent} explicit expressions of the time advancement for geodesic motions for the stated circumstances.

We start with the general static and spherically symmetric spacetime in isotropic coordinates given by

\begin{equation}
ds^{2}=-B(\rho )c_{0}^{2}dt^{2}+A(\rho )\left( d\rho ^{2}+\rho ^{2}d\theta
^{2}+\rho ^{2}\sin ^{2}\theta d\phi ^{2}\right) .
\end{equation}

The post-Newtonian (PN) formalism to some orders [4] is generally used to describe the gravitational theories in a weak gravitational field. This description gives additional advantage of comparing predictions of general relativity with those from any alternative metric theory of gravity. In order to discuss light propagation to any given order, knowledge of every
component of space-time metric to the same order is required [5]. When considered up to the second-PN correction terms, the metric coefficients read 

\begin{equation}
B(\rho )=1-2\frac{\mu }{\rho }+2\beta \frac{\mu ^{2}}{\rho ^{2}}
\end{equation}

and

\begin{equation}
A(\rho )=1+2\gamma \frac{\mu }{\rho }+\frac{3}{2}\delta \frac{\mu ^{2}}{\rho
^{2}}
\end{equation}

where $\beta ,\gamma $ are the Post-Newtonian parameters, $\delta $ can be considered as the second-PN parameter (these parameters are different for different theories [4,6]; in general relativity, all of them are equal to $1$).

Consider that a light signal is transmitted from the surface of the gravitating object (say point $B$ in Fig.4) horizontally (with respect to an observer at $B$) to a nearby point $A$ from where it is reflected back to the point $B$ along the same trajectory. Here B is the point of closest approach for the trajectory. The distance ($\Delta X$) between $A $ and $B$ is very small in comparison to the radius of the gravitating
object. The light signal will travel a null curve of the space-time satisfying $ds^{2}=0$.

\begin{figure}[t]
\centering \includegraphics[width=0.45\textwidth,clip]{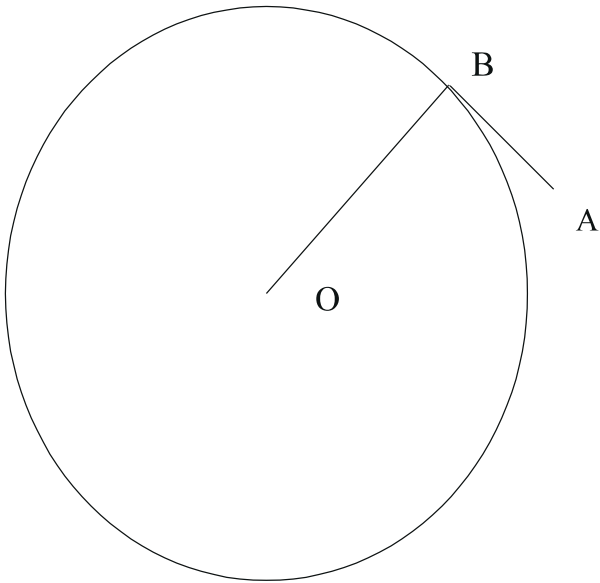} \hfill
\caption{Schematic view of gravitational time advancement situation when
signal travel distance is small}
\end{figure}

To derive time delay to the order of $\mu ^{2}$, one needs to know, to the accuracy of $\mu $, the deviation of photon trajectory from the vertical direction while travelling from $B$ to $A$. The study of geodesic equations reveal that within such accuracy light trajectory does not involve the azimuthal angle and follows straight Euclidean path between $B$ and $A$ for
small $\Delta X$ [5]. Thus, to the second order in $\mu $, the lapse of coordinate time in transiting from $B$ to $A$ is given, after a bit of algebra, by

\begin{equation}
\Delta t_{t}=\Delta L_{BA}\left[ 1+\frac{\mu }{R}+\left( 3/2-\beta \right) 
\frac{\mu ^{2}}{R^{2}}\right] \;,
\end{equation}

where $\Delta L_{BA}$ is the proper distance between $B$ and $A$. Consequently the proper time interval to be measured by the observer at $B$ between transmission and reception of the signal is given by

\begin{equation}
\Delta \tau _{t}=2B^{-1/2}(R)\Delta t_{t}=2\Delta L
\end{equation}

which shows that for motion in the transverse direction there is no gravitational time delay (or advancement) effect at least up to the second PN order when the distance between the points of transmission and reflection is small. The effect of rotation of Earth (as gravitating object) on gravitational time delay for geodesic motion in horizontal direction has been studied in [7] where it was found that rotational contribution is much smaller than the second PN effect.

Next consider the case of radial motion. Restricting to orbits in the equatorial plane ($\theta =\pi /2$), the geodesic equation for $\phi $ leads to

\begin{equation}
\rho ^{2}\frac{d\phi }{dt}=\frac{B}{A}J^{2}
\end{equation}

where $J$ is a constant of integration which in the far field region describes angular momentum of the photon. If initially the motion of the photon is set along the radial direction (so that $J=0$), the above equation warrants the motion would remain radial throughout.

For the PN metric to the leading order the delay in proper time for travel from the point $R,\theta _{0},\phi _{0}$ ($\theta _{0}\equiv \pi /2$ and $\phi _{0}$ are the initial values of the coordinates $\theta $ and $\phi $) to the point $
R^{\prime},\theta _{0},\phi _{0}$ and back is given by

\begin{equation}
\Delta \tau_{\rho } = - 2\mu \left[\left(\frac{R^{\prime}-R}{R} \right) + (\gamma +1) log \frac{R^{\prime}}{R} \right] \;,
\end{equation}

whereas the proper time interval that lapses at $R^{\prime}$ is  

\begin{equation}
\Delta \tau_{\rho } = - 2\mu \left[\left(\frac{R^{\prime}-R}{R^{\prime}} \right) + (\gamma +1) log \frac{R^{\prime}}{R} \right]\;,
\end{equation}

When the propagation distance is small i.e. $R^{\prime}=R+\Delta R$, the total coordinate delay in time for travel would be 

\begin{equation}
\Delta t_{\rho }=2\Delta L\left( 1-\frac{\mu \Delta L}{2R^{2}}\right) \left(
1+\frac{\mu }{R}+(3/2-\beta )\frac{\mu ^{2}}{R^{2}}\right) \;,
\end{equation}

$\Delta L$ corresponds to the proper distance between the points $R$ and $R+\Delta R$. Translating from difference in coordinate time to that in proper time, we have

\begin{equation}
\Delta \tau _{\rho }=B^{-1/2}(\rho )\Delta t_{\rho }=2\Delta L\left( 1-\frac{
\mu \Delta L}{2R_{\oplus }^{2}}\right) \;,
\end{equation}

where we have identified $R=R_{\oplus }$, the error in this identification is small, because of large magnitude of $R_{\oplus }$. Clearly in this case the signal suffers a negative delay due to the negative sign i.e., gravitational time advancement occurs. It is worthwhile to mention that the corrective term follows also from dimensional arguments [8]. Please note
that the calculated effect is independent of PPN $\gamma $ parameter unlike the standard expression for time delay [1,3,4] because we expressed the time delay/advancement results in terms of proper length that includes the $\gamma $ factor (in a gravitational field physical parameters should be expressed in terms of proper quantities particularly in situations like the
present one where proper quantities differ substantially from those of coordinate expressions).

A curious aspect of the above expression is that the time advancement factor is in second order in $1/\rho $ though it is first order in $\mu $. Till now gravitational theories have been tested only to the first order both in $\mu$ and $1/\rho $ in the solar system.

\section{Possibility of experimental detection of time advancement}

The gravitational time advancement due to Earth's gravity can be measured by sending light signal to one of its artificial satellites/space station from where the signal will be reflected back to Earth along the same trajectory and then measuring the total travel time. Note that when a light signal is sent from Earth to one of its artificial satellites/space station, the signal would also come under the influence of Sun's gravity. To overcome this light signal may be sent
(to a satellite) in the perpendicular direction to the axis passing through the Sun and the Earth. In that case the motion of light signal would be in the transverse direction with respect to the Sun and since the distance involved is small, according to the Eq.(8) up to the second PPN order there will be no gravitational time advancement or delay effect due to the Sun. On
the other hand with respect to the Earth the propagation distance is considerable and hence expression given in Eq.(3) will be applicable provided the motion is transverse as in Fig. 1. In that case to the leading order the gravitational time advancement due to Earth's gravity would be

\begin{eqnarray}
c_{0}\Delta \tau _{adv} &=& 2\mu_{\oplus }\frac{\sqrt{R_{sat}^{2} - R_{\oplus}^{2}}}{R_{\oplus}} - 4\mu _{\oplus } \ln \frac{R_{sat}- \sqrt{R_{sat}^{2}-R_{\oplus}^{2}}}{R_{\oplus}}  \nonumber  \\
&&  -  2 \mu _{\oplus }\left( \frac{R_{sat}-R_{\oplus}}{R_{\oplus}+ R_{sat}}\right) ^{1/2} \;,
\end{eqnarray}

where $R_{sat}$ and $R_{\oplus }$ are the coordinate positions of the satellite and the observer at Earth's surface. In the above equation the first term of the right hand side will dominate over the other terms and thus
clearly there will be time advancement. For a high altitude satellite of typical distance $36000$ km, the time of advancement would be about $0.2$ $nsec$ when $\gamma =1$ i.e. for general relativity. In order to measure the time advancement with such a high precision, one has to know the distances with accuracy better than $10$ cm. Instead, in measuring the usual Shapiro effect [3] the distances are treated as unknown parameters and they are determined by fitting the observed times for various positions of reflector. In the proposed case, however, the requirement of high accuracy in distance measurements can be avoided in a better way by repeating the measurement from the satellite i.e. by sending light signal to the Earth from the satellite from where the signal will be reflected back to the satellite and then measuring the total travel time between transmission and reception of the signal. To the leading order the difference in total travel times as measured from Earth and the satellite would be
   
\begin{equation}
\delta \left(c_{0}  \Delta \tau _{adv} \right) \simeq 2\mu _{\oplus}  \sqrt{R_{sat}^{2}-R_{\oplus}^{2}} \left( \frac{1}{R_{\oplus}} - \frac{1}{R_{sat}} \right) \;,
\end{equation}

The magnitude of this difference in travel times would be nearly $0.3$ ns (for $R_{sat} \sim 36000$ km).

On the other hand if the light propagation is radial, the Eqs.(9) and (10) will come to play and the difference in total travel times as measured from Earth and the satellite would be
   
\begin{equation}
\delta \left(c_{0}  \Delta \tau _{adv} \right) \simeq 2\mu _{\oplus } \left(\frac{R_{sat}}{R_{\oplus}} - \frac{R_{\oplus}}{R_{sat}} \right) \;,
\end{equation}

and the magnitude of the difference in travel times would remain nearly the same of that for transverse motion. If a space station is used as reflector, the Eq.(13) will be applicable, as the altitude of a space stations is normally small ($\sim 300$ KM) compare to the radius of Earth.
  
\section{Conclusion}

We conclude the following: Contrary to the common belief gravitational time delay could be negative as well leading to time advancement. This effect arises predominantly because clock runs differently in gravitational field depending on the curvature. The magnitude of the time advancement effect appears nearly half of the difference in gravitational red shift
between the points of reflection and transmission, but it is uncorrelated with the red shift effect. If a light signal is transmitted from a point in a gravitational field and again received at the same point, there will be no gravitational red shift at all i.e. the frequencies of the transmitted and received signal would remain the same in that case.

Light sent away from Earth up to a distance and reflected back along the same path to the point of origin would suffer Earth's gravitational influence. This leads to a negative gravitational time delay that could constitute a potentially new test of general relativity. A possible way of detecting this effect in future is through radar echo like delay experiment with Earth's satellites as reflector.

\end{document}